\documentclass[a4paper,fleqn,usenatbib]{mnras}

\usepackage{newtxtext,newtxmath,multirow}

\usepackage[T1]{fontenc}
\usepackage{ae,aecompl}

\usepackage{graphicx}	
\usepackage{amsmath}	
\usepackage{amssymb}	

\def\gtsima{$\; \buildrel > \over \sim \;$}
\def\ltsima{$\; \buildrel < \over \sim \;$}
\def\gsim{\lower.5ex\hbox{\gtsima}}
\def\lsim{\lower.5ex\hbox{\ltsima}}

\title[Evolution of the reverberation lag in GX 339--4]{Evolution of the reverberation lag in GX 339--4 at the end of an outburst}

\author[B. De Marco et al.]{
B. De Marco$^{1,2}$\thanks{E-mail: bdemarco@camk.edu.pl}
G. Ponti,$^{2}$
P.O. Petrucci,$^{3}$
M. Clavel,$^{4}$
S. Corbel,$^{5,6}$
R. Belmont$^{7,8}$,
\newauthor S. Chakravorty$^{9}$,
M. Coriat$^{7,8}$,
S. Drappeau$^{7,8}$,
J. Ferreira$^{3}$,
G. Henri$^{3}$,
J. Malzac$^{7,8}$,
\newauthor J. Rodriguez$^{5}$,
J. A. Tomsick$^{4}$,
F. Ursini$^{3,10}$,
A. A. Zdziarski$^{1}$
\\
$^{1}$Nicolaus Copernicus Astronomical Center, Polish Academy of Sciences, Bartycka 18, PL-00-716 Warsaw, Poland\\
$^{2}$Max-Planck-Institut f\"ur Extraterrestrische Physik, Giessenbachstrasse 1, D-85748 Garching, Germany\\
$^{3}$Universit\'e Grenoble Alpes, CNRS, IPAG, F-38000 Grenoble, France\\
$^{4}$Space Sciences Laboratory, 7 Gauss Way, University of California, Berkeley, CA 94720, USA\\
$^{5}$Laboratoire AIM (CEA/IRFU-CNRS/INSU-Universit\'e Paris Diderot), CEA DRF/IRFU/DAp, F-91191 Gif-sur-Yvette, France\\
$^{6}$Station de Radioastronomie de Nan\c{c}ay, Observatoire de Paris, PSL Research University, CNRS, Univ. Orle\'ans, 18330 Nan\c{c}ay, France\\
$^{7}$Universit\'e de Toulouse; UPS-OMP; IRAP; Toulouse, France\\
$^{8}$CNRS; IRAP; 9 Av. colonel Roche, BP 44346, F-31028 Toulouse cedex 4, France\\
$^{9}$Department of Physics, Indian Institute of Science, Bangalore, 560012, India\\
$^{10}$Dipartimento di Matematica e Fisica, Universit\`a degli Studi Roma Tre, via della Vasca Navale 84, 00146, Roma, Italy
}

\date{Accepted XXX. Received YYY; in original form ZZZ}

\pubyear{2016}

\begin{document}
\label{firstpage}
\pagerange{\pageref{firstpage}--\pageref{lastpage}}
\maketitle

\begin{abstract}
We studied X-ray reverberation lags in the BHXRB GX 339--4 at the end of the 2014-2015 outburst. We analysed data from a {\it XMM-Newton} campaign covering the end of the transition from the soft to the hard state, and the decrease of luminosity in the hard state. During all the observations we detected, at high frequencies, significant disc variability, responding to variations of the power law emission with an average time delay of $\sim0.009\pm 0.002$ s. These new detections of disc thermal reverberation add to those previously obtained and suggest the lag to be always present in hard and hard-intermediate states. Our study reveals a net decrease of lag amplitude as a function of luminosity. We ascribe this trend to variations of the inner flow geometry. A possible scenario implies a \emph{decrease} of the inner disc truncation radius as the luminosity increases at the beginning of the outburst, followed by an \emph{increase} of the inner disc truncation radius as the luminosity decreases at the end of the outburst. Finally, we found hints of FeK reverberation ($\sim3\sigma$ significance) during the best quality observation of the {\it XMM} monitoring. The lag at the FeK energy has similar amplitude as that of the thermally reprocessed component, as expected if the same irradiated region of the disc is responsible for producing both the thermalized and reflected component. This finding suggests FeK reverberation in BHXRBs to be at the reach of current detectors provided observations of sufficiently long exposure are available.
\end{abstract}




\section{Introduction}
\label{sec:intro}
Black hole X-ray binaries (BHXRB) in outburst display hysteresis between Compton-dominated and disc-dominated accretion regimes (\emph{hard} and \emph{soft states}, respectively; e.g. Miyamoto et al. 1995). Transitions occur fast, on typical time scales of a few days (e.g. Dunn et al. 2010), throughout the so-called \emph{intermediate state} (e.g. Mend\'ez \& van der Klis 1997). Given the dramatic variations of spectral and timing properties of the source (e.g. Nowak 1995; Fender et al. 2004; Homan \& Belloni et al. 2005; Belloni et al. 2005; Dunn et al. 2010; Mu\~noz-Darias et al. 2011), these transitions are thought to be related to major changes in the properties of the inner accretion flow (e.g. Done, Gierli\'nski \& Kubota 2007). The adopted paradigm to explain the observed phenomenology is that of a geometrically thin, optically thick accretion disc with an inner truncation radius varying as a function of the accretion rate of the source (e.g. Esin et al. 1997). In this scenario, the \emph{soft state} is energetically dominated by a standard, radiatively efficient accretion disc extending down to the innermost stable orbit (ISCO) of the BH (Shakura \& Sunyaev 1973; Novikov \& Thorne 1973). On the other hand, when the source is in the low luminosity \emph{hard state} the disc recedes and is replaced by a radiatively inefficient, optically thin plasma in the inner parts (the so-called ``hot flow'', responsible for Compton up scattering the disc thermal photons to hard X-ray energies; e.g. Narayan \& Yi 1995). 

Changes in the accretion disc structure offer a theoretical framework to explain the entire outburst evolution. Several X-ray timing properties (e.g. Done, G\'ierlinski \& Kubota 2007), most notably the appearance of type-C quasi-periodic oscillation (QPO; e.g. Remillard et al. 2002) and the increase of their frequency as the source spectrum softens (e.g. Rodriguez et al. 2004), agree with a scenario of an evolving disc truncation radius (e.g. Stella \& Vietri 1998; Rodriguez et al. 2002; Ingram et al. 2009). However, more direct observational evidences appear controversial. Indeed, spectral fits support the existence of a standard disc reaching the ISCO during disc-dominated states (e.g., Gierli\'nski \& Done 2004; Steiner et al. 2010; Plant et al. 2014), and a highly truncated one in quiescent/low-luminosity states (e.g. Tomsick et al. 2009). However, the extent to which the inner disc is truncated during high luminosity \emph{hard} and \emph{intermediate} states is unclear. Previous measurements of disc truncation have mostly relied on spectral modelling of the reflection spectrum resulting from the reprocessing of the Comptonized hard X-ray photons in the accretion disc (e.g. Guilbert \& Rees 1989). Nonetheless, reported results reveal several discrepancies (e.g. Miller et al. 2004; Miller, Homan \& Miniutti 2006a; Miller et al. 2006b; Kolehmainen et al. 2014; Petrucci et al. 2014; F\"urst et al. 2015; Plant et al. 2015; Garc\'ia et al. 2015; Basak \& Zdziarski 2016).

In De Marco et al. (2015) and De Marco \& Ponti (2016) we have used the alternative approach of studying the delayed response of the dense material of the disc to variations in the illuminating hard X-ray flux (Blandford \& McKee 1982). These ``X-ray reverberation lags'' map the relative distance between the disc and the Comptonizing region, so that they can be used to constrain the geometry of the inner accretion flow (Uttley et al. 2014).
The first detection of a X-ray reverberation lag in a BHXRB was reported by Uttley et al. (2011) during one \emph{hard state} observation of GX 339--4. The lag, detected in the soft X-ray band, was interpreted as thermal reprocessing of the hard X-ray photons in the disc. Extending this analysis to a larger sample, we found evidence of a trend of decreasing thermal reverberation lag amplitude as a function of luminosity throughout the \emph{hard state} at the beginning of the outburst in GX 339--4 and H1743--322 (De Marco et al. 2015; De Marco \& Ponti 2016). This trend is in agreement with predictions of truncated disc models (e.g. Esin et al. 1997). Moreover, even when the source is in a very bright \emph{hard state} (i.e. at Eddington-scaled luminosities of $\rm{L_{3-10 keV}/L_{Edd}}\sim$0.02) the reverberation lag amplitude is still large ($\sim 0.003\ {\rm s}$; De Marco et al. 2015), mapping distances of a few tens of gravitational radii $r_g$ (assuming a BH mass of $\sim 10\ \rm{M_{\odot}}$), possibly indicating substantial disc truncation throughout the entire \emph{hard state} (unless the primary X-ray source is located at significant distance above the accretion disc, e.g. in a jet-like configuration, Fabian et al. 2014).

In this paper we present the analysis of X-ray reverberation lags in GX 339--4 during the final phases of its last outburst. 
GX 339--4 is one of the most active BHXRBs, characterized by a fairly regular activity, with many (about 19; Corral-Santana et al. 2016) X-ray outbursts being observed since its discovery (Markert et al. 1973). The most recent outburst started in 2014 October (Yan et al. 2014), and lasted for about one year. Currently, only secure lower limits on the mass of the BH in GX 339--4 exist ($\rm{M_{BH}}\geq 6 \rm{M_\odot}$ for an orbital inclination angle $i\lsim80^\circ$; Mu\~noz-Darias, Casares \& Mart\'inez-Pais 2008). The source is well-known for having shown all the known accretion states (e.g. Belloni et al. 2005), therefore it is an ideal target for studying the geometrical distribution of the inner accretion flow as a function of the accretion state. 
The data analysed in this paper are from a {\it XMM-Newton} monitoring campaign (carried out in 2015 August-September) triggered during the transition from the \emph{soft to the hard state}. 

The paper is organized as follows. In Sect. \ref{sec:reduction} we report details about the data reduction procedure. In Sect. \ref{sec:state} we make use of X-ray spectral and timing diagnostics to determine the accretion state of the source during each observation. In Sect. \ref{sec:specs} we perform a broad band X-ray spectral analysis to determine the relative contribution of the disc and Comptonization components. In Sect. \ref{sec:lags} we report on the analysis of X-ray lags. In Sect. \ref{sec:results} we study the evolution of the reverberation lag as a function of luminosity and carry out a comparison with results from previous analysis (De Marco et al. 2015). Results are discussed in Sect. \ref{sec:discussion}.

\section{Data reduction}
\label{sec:reduction}
The data set analysed in this paper consists of 6 {\it XMM-Newton} observations carried out between 2015 August 28th and September 30th.  Table \ref{table1} reports the log of the observations. For brevity we will use the nomenclature reported in this table to refer to each of them.

For our analysis we used data from the EPIC-pn detector (Str\"uder et al. 2001) given its large effective area and high time resolution. This instrument was operated in Timing mode during the first four observations (O1 to O4), and in SmallWindow mode during the last two (O5 and O6).
The data reduction was carried out following standard procedures, using the {\it XMM} Science Analysis System (SAS v15) and calibration files as of August 2016. 
The exposures range between $\sim 15$ ks and $\sim$36 ks.

Background flares do not affect these observations, apart from a single flare during O1. Removal of this flare reduces the exposure from $\sim$15 ks to $\sim$6 ks. The main effect of this flare is to slightly harden the X-ray continuum ($\Delta \Gamma\sim$0.13 after fitting with a power law model). Therefore, for our spectral analysis we filtered this flare out. However, this flare does not affect the analysis of X-ray lags (we verified that results obtained excluding the flare are consistent). Therefore, in order to have a sufficiently long exposure and maximize the signal-to-noise of lag measurements we did not perform the filtering for the X-ray timing analysis.

EPIC-pn source counts were extracted in the interval RAWX: 28-48 for the observations in Timing mode (this extraction area corresponds to an angular size of $\sim$86 arcsec along RAWX), and in a circular region of 40 arcsec radius for the observations in SmallWindow mode. 
Using the task \emph{epatplot} we verified the presence of pile-up in the data. Deviations of the pattern-fraction distribution from the theoretical curves are mild and below $\sim 1$ percent.

We extracted spectra of the source using the tasks \emph{rmfgen} and \emph{arfgen} to generate response matrices. Background spectra have been extracted from two rectangular regions not contaminated by the source during the observations in SmallWindow mode (O5 and O6). In Timing mode it is not possible to isolate source-free regions. Therefore, for these observations we used the background spectrum obtained during O5, after checking the background counts are properly rescaled for the area of the source counts extraction region.
 
Finally, we verified that the background does not have significant effects on the analysis of X-ray lags (Sects. \ref{sec:lagfreq} and \ref{sec:lagE}). Therefore, in order to retain a sufficiently high signal-to-noise ratio, background subtraction was not performed for the X-ray spectral-timing analysis.

\section{Accretion state of GX 339--4 during the {\it XMM} monitoring}
\label{sec:state}
\begin{figure}
	\includegraphics[width=\columnwidth]{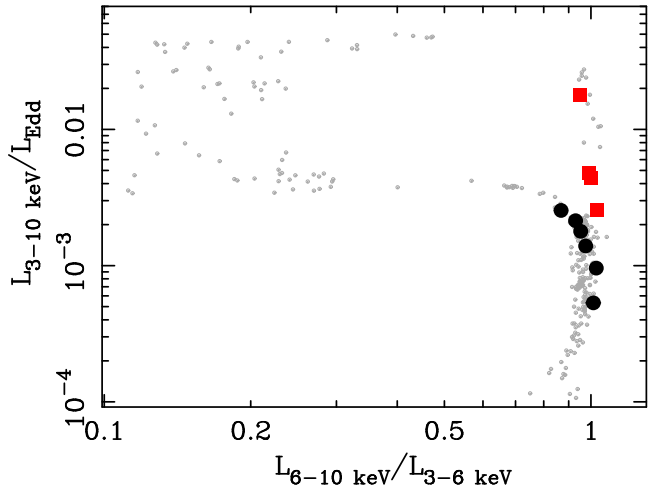}
    \caption{The HID of GX339--4. The black dots refer to the {\it XMM} observation analysed in this paper. The red squares are the archival {\it XMM} observations of GX 339--4 analysed in De Marco et al. (2015), and performed during the 2009 and 2004 outbursts. For reference, the gray dots show the pattern of a full outburst observed by {\it RXTE} in 2009. A $\sim 10$ correction factor should be applied in order to convert the reported $3-10\ {\rm keV}$ luminosities into bolometric luminosities.}
    \label{fig:HID}
\end{figure}

To determine the accretion state of the source during the analysed observations we inferred the position of the source on the hardness-intensity diagram, HID (e.g. Homan \& Belloni et al. 2005). We simultaneously fitted the spectrum of all the observations with a simple model for the 3-10 keV continuum. The model includes a power law absorbed by a Galactic column of cold gas ($Tbabs*powerlaw$ in Xspec; Wilms, Allen \& McCray 2000). To avoid complexities associated with the Fe K line complex we discarded the energy range 5-8 keV. The column density parameter of the cold absorption component was fixed at the value $N_{\rm H}=6\times10^{21} cm^{-2}$ as reported in Galactic HI surveys (e.g. Dickey \& Lockman 1990) and in agreement with spectral fits of archival data sets (e.g. De Marco et al. 2015). Table \ref{table1} lists the best-fit values of the photon index and the 3-6 keV and 6-10 keV flux. The corresponding HID is shown in Fig. \ref{fig:HID}. The black dots refer to the {\it XMM} monitoring. For a comparison we also plot (red squares) the position in the HID of the archival {\it XMM} observations analysed in De Marco et al. (2015), re-analysed here using updated calibration files. 
Eddington-scaled luminosities were computed adopting a distance of $8$ kpc (Zdziarski et al. 2004) and a BH mass of $12\ \rm{M_\odot}$\footnote{Note that here we assume a slightly higher value of BH mass than in De Marco et al. (2015). Indeed, using the orbital parameters derived in Hynes et al. (2003) and correcting for the radial velocity of the companion star as in Mu\~noz-Darias et al. (2008), values of $\rm{M}<12\ \rm{M_\odot}$ would imply a binary inclination angle $i\gsim60^\circ$. However, this conflicts with the observational evidences that GX 339--4 is a low-inclination system (e.g. Ponti et al. 2012a, 2016; Mu\~noz-Darias et al. 2013) unless the inner part of the disc and the binary orbit are misaligned (e.g. Maccarone 2002).} (in good agreement with recent estimates; Parker et al. 2016). 
Note that the uncertainty on the distance and the BH mass affect the normalization of the HID. Assuming a 20 percent uncertainty on both parameters we derive a factor $\sim$2 uncertainty on the normalization of the HID. For the reported $3-10\ {\rm keV}$ luminosities to be converted into bolometric luminosities we estimate a $\sim 10$ correction factor (i.e. assuming the broad band model that will be described in Sect. \ref{sec:specs}).

In order to better visualize which phases of the outburst are spanned by the {\it XMM} observations we overplot the HID of GX 339--4 during its 2009 outburst as determined using PCA {\it Rossi X-ray Timing Explorer} ({\it RXTE}) archival observations (light gray small dots).

\begin{table*}
	\centering
	\caption{Log of the observations of GX 339--4 analysed in this paper. The table reports: (1) the {\it XMM} observation ID; (2) the observation date; (3)  the nomenclature used throughout the paper to refer to each observation; (4) the EPIC pn observing mode; (5) the net exposure time used for the X-ray timing analysis (note that this is the same as used for the spectral analysis with the exception of O1, see Sect. \ref{sec:reduction} for details); (6) the best-fit slope parameter after fitting the 3-10 keV continuum with an absorbed power law model (see Sect. \ref{sec:state} for details); (7) the 3-6 keV source flux; (8) the 6-10 keV source flux; (9) the 3-10 keV $F_{\rm var}$ (see Sect. \ref{sec:state} for more details) obtained integrating over the frequency range 0.05-60 Hz.}
			\label{table1}
	\begin{tabular}{lcccccccc} 
		\hline
		   (1)    & (2) & (3)    & (4)                  &  (5)              &     (6)                       &  (7)                            & (8)            & (9)       \\
		  ID &   Date   &Obs & Mode & Exposure        & $\Gamma$ & $F_{3-6\ {\rm keV}}$ & $F_{6-10\ {\rm keV}}$ & $F_{\rm var}$ \\
		  	&  [yyyy-mm-dd] &    &            &    [ks]              &                     & [$10^{-10}\ {\rm erg/s/cm^{-2}}$] & [$10^{-10}\ {\rm erg/s/cm^{-2}}$] &   \\
		\hline
		0760646201 & 2015-08-28 & O1 & Timing &  14.9 & 1.77$\pm$0.02 & 2.59$\pm$0.06 & 2.25$\pm$0.05 & $0.28\pm0.01$\\
		0760646301 & 2015-09-02 & O2  & Timing & 15.7  & 1.66$\pm$0.01 & 2.11$\pm$0.03 & 1.96$\pm$0.03 & $0.32\pm0.01$\\
		0760646401 & 2015-09-07 & O3 &Timing & 20.1 & 1.62$\pm$0.01   & 1.74$\pm$0.03 & 1.66$\pm$0.03 & $0.35\pm0.01$ \\
	      0760646501 & 2015-09-12 & O4 & Timing & 18.6  & 1.58$\pm$0.01 & 1.34$\pm$0.02 & 1.31$\pm$0.02 & $0.37\pm0.01$ \\
		0760646601 & 2015-09-17 & O5 & SmallWindow & 36.5  & 1.49$\pm$0.01 & 0.90$\pm$0.01 & 0.93$\pm$0.01 & $0.32\pm0.01$ \\
		0760646701 & 2015-09-30 & O6 & SmallWindow & 33.4 & 1.51$\pm$0.01 & 0.50$\pm$0.01 & 0.51$\pm$0.01  & $0.36\pm0.02$\\
		\hline
	\end{tabular}
\end{table*}

The {\it XMM-Newton} monitoring of GX 339--4 covers the very last stages of the \emph{soft-to-hard state} transition (observations O1 and O2) and the decrease of luminosity through the \emph{hard state} preceding the return to quiescence (O3-to-O6). The source undergoes a factor $\sim5$ decrease of 3-10 keV luminosity between the first and the last observation.

\begin{figure*}
	\includegraphics[width=0.8\textwidth]{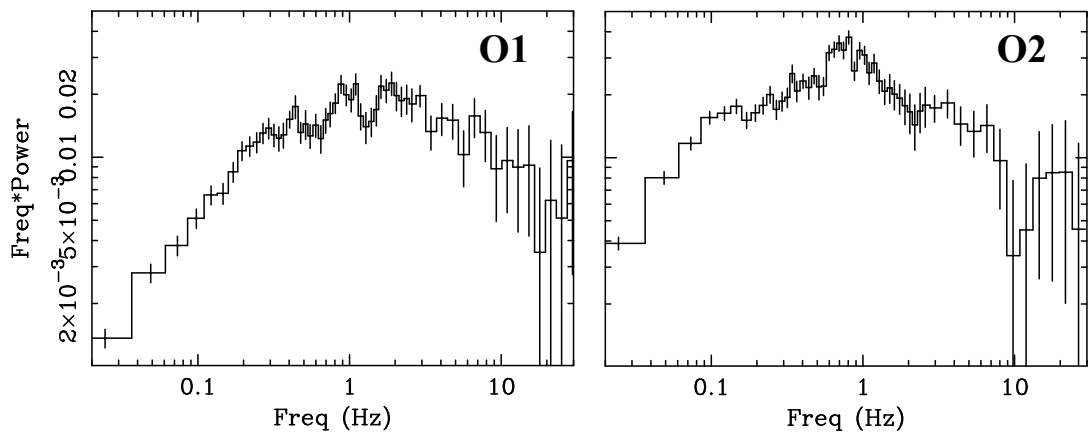}
    \caption{The power spectrum of GX339--4 during O1 (\emph{left panel}) and O2 (\emph{right panel}), in the energy band 3-10 keV. The power spectra are Poisson noise-subtracted and displayed adopting the fractional root-mean-square normalization (Miyamoto et al. 1991).}
    \label{fig:PSD}
\end{figure*}

Also the X-ray timing properties of the source are typical of these phases of the outburst. 
The power spectrum of the source is flat-topped (Fig. \ref{fig:PSD} shows the 3-10 keV power spectrum of O1 and O2), as commonly observed after the \emph{soft-to-hard} transition and during the \emph{hard state} (e.g. Belloni et al. 2005). Since type-C QPOs are occasionally observed in association with flat-topped broad band noise (e.g. Casella, Belloni \& Stella 2005; Belloni et al. 2005; F\"urst et al. 2016), we checked for their presence fitting the power spectra with Lorentzian components. We found the analysed data set to be dominated by broad-band noise components, without any clear evidence of the presence of QPOs. One narrower, QPO-like, noise component is observed only during O1 and O2 (Fig. \ref{fig:PSD} ), peaking at frequencies $\sim 0.9\ {\rm Hz}$ and $0.7\ {\rm Hz}$, respectively. However, the corresponding quality factor (defined as $\nu/FWHM$, e.g. Belloni, Psaltis \& van der Klis 2002) is too low (respectively $\sim3$ and $\sim1.6$) for this component to be associated with a type-C QPO\footnote{It is worth noting that such low values of the quality factor are characteristic of type-A QPOs. Nonetheless, these typically appear when the source moves from the \emph{hard} to the \emph{soft state}, have higher peak frequencies and are associated with a weak, red-noise continuum (e.g. Casella, Belloni \& Stella 2005; Belloni et al. 2005).}.

Finally, we computed the fractional root-mean-square variability amplitude ($F_{\rm var}$, measuring the fraction of intrinsically variable emission with respect to the total emission; e.g. Nandra et al. 1997; Vaughan et al. 2003; Ponti et al. 2012b) in the energy range 3-10 keV and over the frequency range 0.05-60 Hz. The resulting estimates are listed in Table \ref{table1}. During the monitoring, GX 339--4 is highly variable, with $F_{\rm var}$ ranging between $\sim$28-37 percent and showing an increasing trend as the outburst proceeds. This is consistent with the source exiting the \emph{soft-to-hard} transition branch and returning to the \emph{hard state} (e.g. Mu\~noz-Darias et al. 2011).

\section{Spectral analysis}
\label{sec:specs}
Detailed spectral analysis of this data set is deferred to an upcoming paper (Petrucci et al. in preparation). Here we are interested in estimating the relative contribution of the disc component (which should contain contribution from thermal reprocessing) relative to the Comptonization component. These estimates will also be used to define the optimal energy bands for the analysis of X-ray lags (Sect. \ref{sec:lagfreq}).

We carried out a fit using a simple model for the broad band continuum. The model includes: a multi-temperature black body component ($diskbb$ in Xspec, e.g. Mitsuda et al. 1984) to account for thermal disc emission; a Comptonization component ($nthComp$ in Xspec, Zdziarski, Johnson \& Magdziarz 1996; \.{Z}ycki, Done \& Smith 1999), to account for the primary hard X-ray emission. The high-energy cut-off of the Comptonization component was fixed at the value of 100 keV and the seed photon temperature tied to the inner disc temperature. Both components are absorbed by a column of cold gas ($Tbabs$ in Xspec) that we fixed at the value $N_{\rm H}=6\times 10^{21} \rm{cm^{-2}}$ (note that leaving this parameter free to vary yields $\sim10$ percent smaller $N_{\rm H}$ values and does not change significantly our results).

\begin{figure*}
	\includegraphics[width=0.8\textwidth]{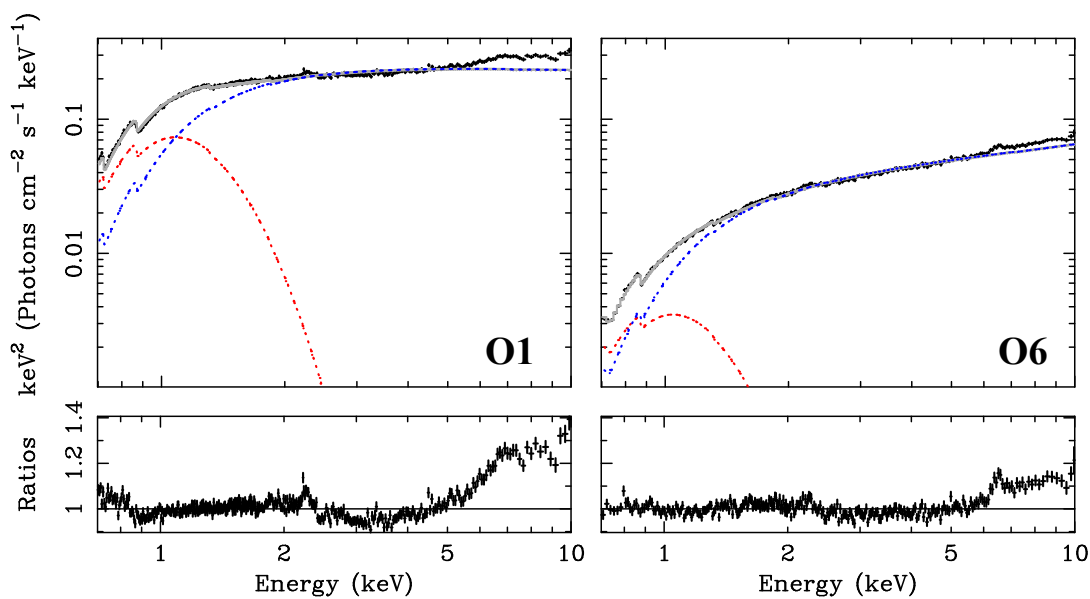}
    \caption{The plots show the EPIC pn unfolded spectra of the first (O1, \emph{left panel}) and the last observation (O6, \emph{right panel}) of the monitoring. Overplotted is the best-fit model (solid gray curve) and the single components of the model (i.e. $diskbb$, red dotted curve, and $nthComp$, blue dotted curve), as obtained excluding the energy range $5-8$ keV, dominated by the FeK complex. Bottom panels show the ratios to the best-fit model.}     
\label{fig:spectra}
\end{figure*}

The spectra of the different observations were fit simultaneously in the energy band 0.7-10 keV. The chosen low-energy cut allows us to avoid distortions on the pattern fraction distribution due to electronic noise in Timing mode data (i.e. O1-to-O4, as recommended in Guainazzi, Haberl \& Saxton 2010). For consistency we used the same energy range for the fit of data acquired in SmallWindow mode (O5 and O6). However, we checked that
the results of the fits do not change significantly when extending the analysis down to 0.3 keV.
Fig. \ref{fig:spectra} shows the (unfolded) spectra of GX 339--4 at the beginning (O1) and at the end of the monitoring (O6) together with the corresponding best-fit models. The fits have been carried out excluding the energy range 5-8 keV, in order to avoid complexities associated with the Fe K complex. This component clearly dominates the hard X-ray part of the spectrum and requires a treatment with self-consistent reflection models (Petrucci et al. in preparation) though we checked that accounting for this component using a simple gaussian does not have significant impact on the results reported below.

The simultaneous fit of all the observations yields $\chi^2/dof=14406/7549$. We verified that this is mostly due to the strong residuals at the energies close to the $\sim1.8\ {\rm keV}$ and $\sim 2.2\ {\rm keV}$ edges of the response matrix, particularly during the observations in Timing mode. We ascribe such residuals to the $\sim 20\ {\rm eV}$ accuracy of the energy scale reconstruction (e.g. Guainazzi 2013; 2014; see also discussion in Kolehmainen et al. 2014), as the shift we observe between the data and the model is of this same order. After discarding the energy range 1.5-2.5 keV the fit considerably improves ($\chi^2/dof=8594/6337$; hereafter we report best-fit values obtained by excluding this energy range). Finally, we note that the excess residuals observed at energies $\lsim 0.9$ keV during O1 have been previously observed in GX 339-4 (Kolehmainen et al. 2014), and might indicate the need for a more complex model for the soft component. Table \ref{table2} reports the best-fit values for the relevant parameters of the model.

 For all the spectra, the disc thermal component is required to fit the soft X-rays. The best-fit inner disc temperature is in the range $kT_{d} \sim 0.18-0.20\ {\rm keV}$. We observe a mild decrease of the disc temperature between the first and the last observations. However, our data are not very sensitive to this parameter because of the soft band sensitivity limit of the instrument, which allows us to cover only the hard tail of the disc thermal component.
 
\begin{figure}
	\includegraphics[width=\columnwidth]{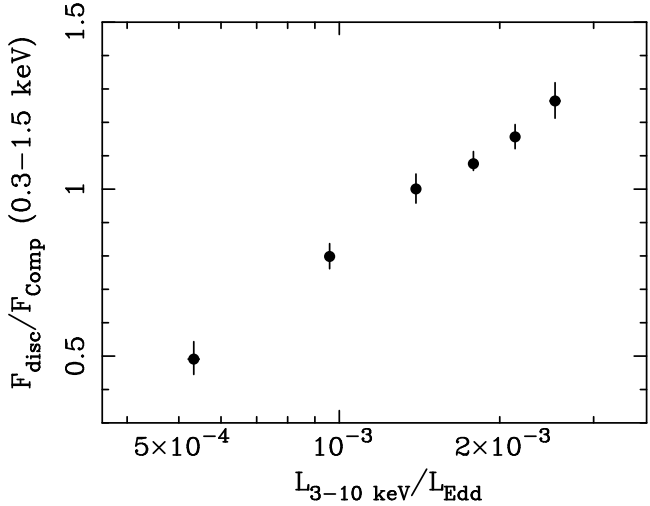}
    \caption{Disc-to-Comptonized flux ratios (in the energy range 0.3-1.5 keV) as a function of 3-10 keV Eddington-scaled luminosity. The fluxes are computed using the best-fit model described in Sect. \ref{sec:specs} and without correction for cold absorption.}
    \label{fig:ratios}
\end{figure}

We estimated the relative contribution of the disc thermal component in terms of disc-to-Comptonized flux ratios in the energy range 0.3-1.5 keV (the best-fit models have been extrapolated from 0.7 keV down to 0.3 keV). In Fig. \ref{fig:ratios} we plot this ratio as a function of 3-10 keV Eddington-scaled luminosity. 
Since we are interested in the fraction of observed disc photons relative to Comptonized photons we report the flux ratios for the absorbed spectrum (the intrinsic, unabsorbed, disc-to-Comptonized flux ratios are higher by a factor $\sim3$).
The disc relative contribution in the soft band is always relatively high ($F_{\rm disc}/F_{\rm Comp}\gsim0.5$), despite the net decrease of disc luminosity (by a factor $\sim$20 in the $0.3-1.5\ {\rm keV}$ band between O1 and O6) as the outburst proceeds.

\section{X-ray lags analysis}
\label{sec:lags}
\subsection{Lag-frequency spectra}
\label{sec:lagfreq}
We first studied the X-ray lags as a function of the Fourier frequency. These allow us to investigate the causal connection between the hard X-ray primary continuum and the disc component by measuring the distribution of lag amplitude over frequencies corresponding to different time scales of variability. 
Using our best-fit spectral models (Sect. \ref{sec:specs}) we defined three energy bands, hereafter referred to as \emph{very soft} (0.3-0.7 keV), \emph{soft} (0.7-1.5 keV), and \emph{hard} (3-5 keV). These bands pinpoint, respectively, the maximum disc contribution (disc-to-Comptonized flux ratio $\sim2-4$), the drop of disc flux with respect to the Comptonized flux occurring at $\sim 1\ {\rm keV}$ (disc-to-Comptonized flux ratio $\sim0.3-1$), and the hard X-ray primary emission-dominated band (excluding contribution from the FeK line component), where disc contribution is negligible. For each observation, Table \ref{table2} lists the disc-to-Comptonized flux ratios in the chosen \emph{very soft} and \emph{soft} bands. Note that, despite the fact that the 0.3-0.7 keV band might be more affected by higher background due to low-energy electronic noise in Timing mode data (e.g. Guainazzi, Haberl \& Saxton 2010), this is not expected to produce any spurious effect on our lag measurements, apart from a possible decrease of the signal-to-noise.

\begin{figure}
	\includegraphics[width=0.93\columnwidth]{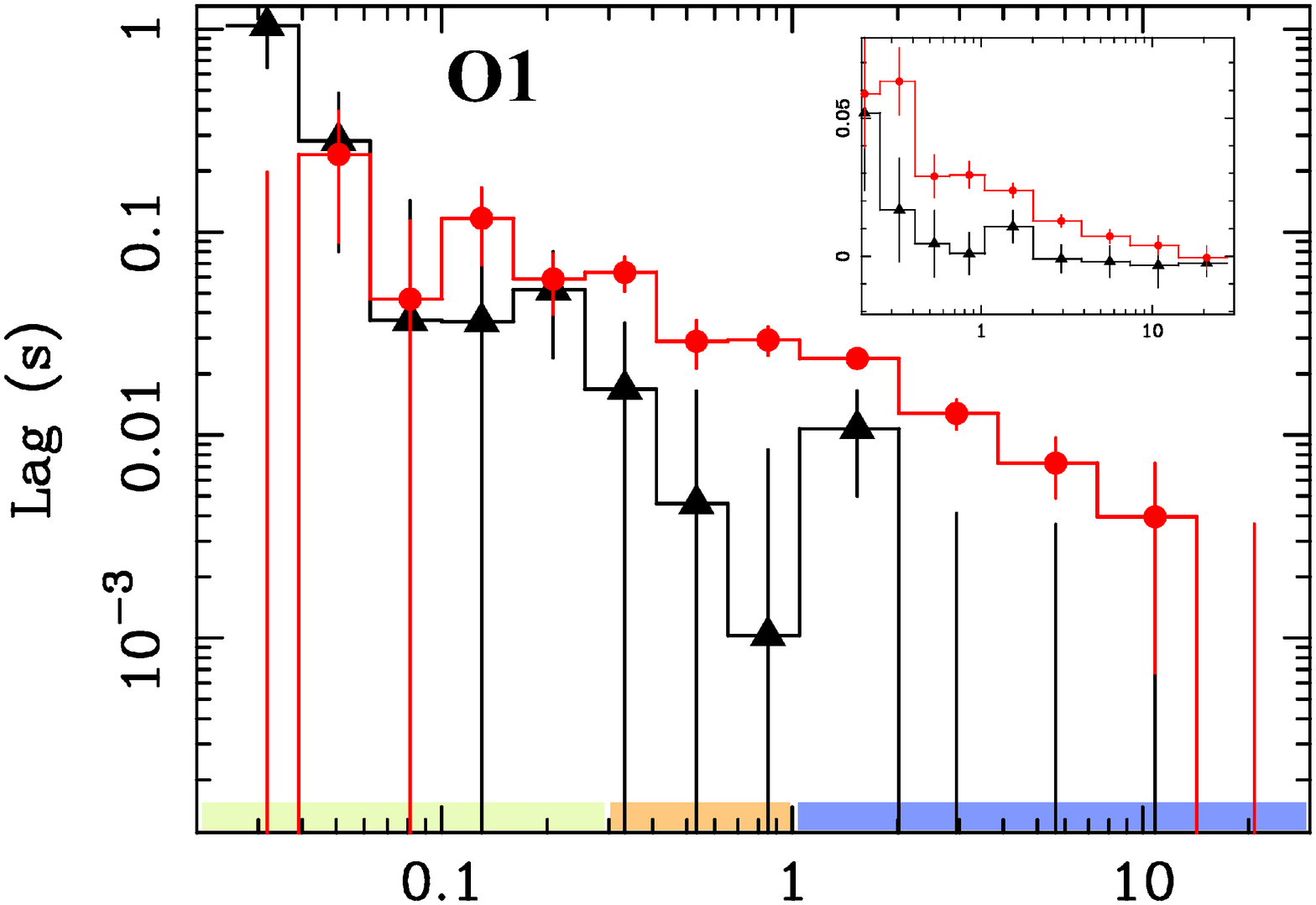}
	\vspace{0.2cm}
	\includegraphics[width=0.93\columnwidth]{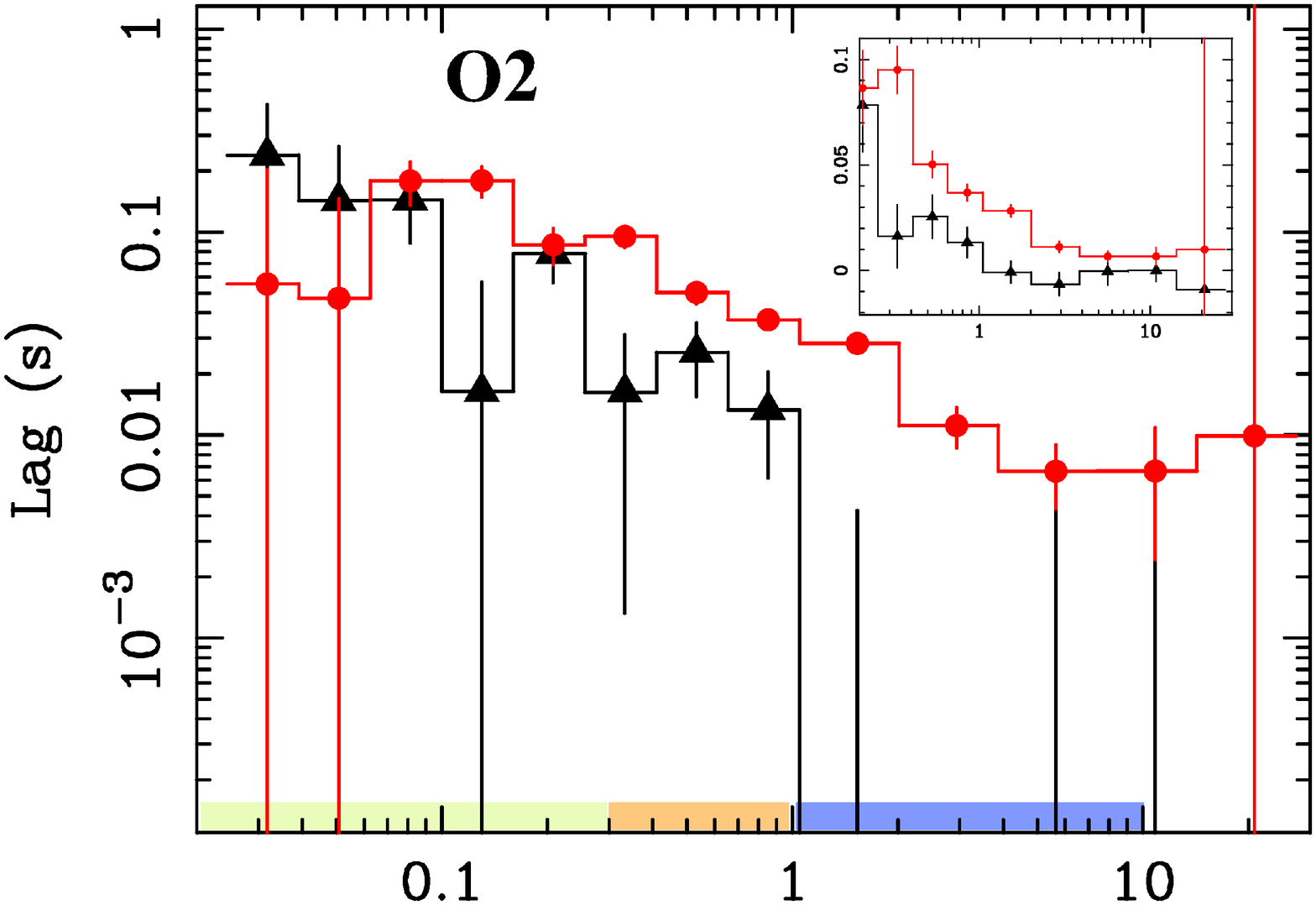}
	\vspace{0.2cm}
	\includegraphics[width=0.93\columnwidth]{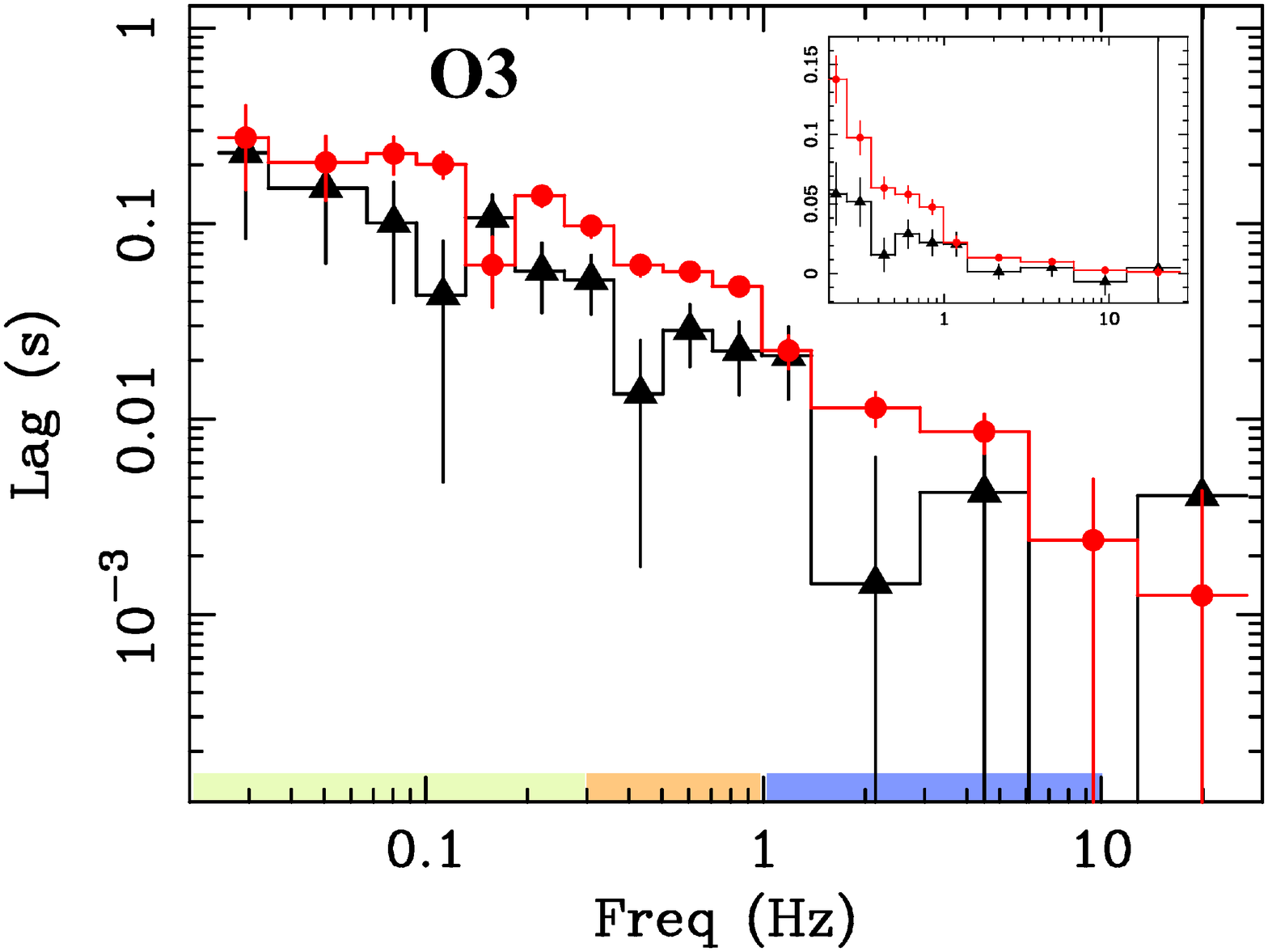}

    \caption{\emph{Very soft}-\emph{soft} (black triangles) and the \emph{soft}-\emph{hard} (red dots) lag-frequency spectra of observations O1 (\emph{top panel}), O2 (\emph{middle panel}), and O3 (\emph{lower panel}). Coloured bars on the x-axis mark the \emph{low}, \emph{medium}, and \emph{high frequency} intervals chosen for the extraction of lag-energy spectra. The insets show a zoom of the 0.2-30 Hz frequency range (where the drop of \emph{very soft}-\emph{soft} lag amplitude is observed) using a linear scale for the y-axis.}
    \label{fig:lagfreqO1O2}
\end{figure}

\begin{table}
	\centering
	\caption{The table reports relevant information derived from the spectral analysis of EPIC pn data: the best-fit slope (column 2) and inner disc temperature (column 3) as obtained from the fit of the EPIC pn spectra with the $Tbabs*(diskbb+nthComp)$ model in Xspec; the disc-to-Comptonized flux ratio (column 4) as obtained from the same fits, in the \emph{very soft} (0.3-0.7 keV) and \emph{soft} (0.7-1.5 keV) bands.}
			\label{table2}
	\begin{tabular}{ccccc} 
		\hline
		  (1)    &  (2)             & (3)            &  \multicolumn{2}{c}{(4)}                                                \\
		   Obs & $\Gamma$ & $k T_{d}$  & \multicolumn{2}{c}{$F_{\rm{disc}}$/$F_{\rm{Comp}}$}  \\
		          &                   &                 & \emph{very soft}                            & \emph{soft}             \\
		          &                   &   [keV]     &   \emph{band}                          &       \emph{band}            \\
		\hline
		O1 & 1.91$\pm$0.01  &0.192$\pm$0.001  &   4.48$\pm0.20$   & 1.17$\pm0.05$ \\
		O2 & 1.73$\pm$0.01  &0.201$\pm$0.001  &    4.14$\pm0.13$   & 1.08$\pm0.03$ \\
		O3 & 1.68$\pm$0.01  &0.199$\pm$0.001  &    4.01$\pm0.14$   & 1.00$\pm0.03$ \\
		O4 & 1.64$\pm$0.01  &0.192$\pm$0.001  &    4.01$\pm0.18$   & 0.92$\pm0.04$ \\
		O5 & 1.59$\pm$0.01  &0.180$\pm$0.001  &    3.73$\pm0.18$   & 0.73$\pm0.04$ \\
		O6 & 1.59$\pm$0.01  &0.180$\pm$0.003  &    2.29$\pm0.24$  & 0.45$\pm0.05$ \\
		\hline
	\end{tabular}
\end{table}
Using standard cross-spectral analysis techniques (e.g. Nowak et al. 1999; Uttley et al. 2014) we computed the lag-frequency spectra of all the six observations, between the \emph{very soft} and the \emph{soft} band, and between the \emph{soft} and \emph{hard} band. 
To this aim we extracted EPIC pn light curves in each of the selected energy bands and with a time resolution of 6 ms. The light curves were then divided into segments of $\sim$ 41 s length (therefore allowing us to cover the frequency range $\sim0.024-83\ {\rm Hz}$), and the cross-spectrum (e.g. equation 7 of Uttley et al. 2014) computed for each segment. We then averaged over the cross-spectra from the different segments to obtain an estimate of the time lag as a function of frequency. In Fig. \ref{fig:lagfreqO1O2} we show the lag-frequency spectra of the highest count rate observations, O1, O2, and O3 (a positive lag amplitude corresponds to a hard lag). In Fig. \ref{fig:lagfreqO1O2} the \emph{very soft}-\emph{soft} (black triangles) and the \emph{soft}-\emph{hard} (red dots) lag-frequency spectra are overplotted. 

Past studies of X-ray lags in BHXRBs (mostly based on the analysis of {\it RXTE} data) revealed the presence of ubiquitous hard lags (a delay of the higher-energy bands with respect to the lower-energy bands) associated with the primary hard X-ray continuum (e.g. Nowak et al. 1999; Pottschmidt et al. 2000). Their properties can be explained by models invoking inward propagation of mass accretion rate fluctuations in the accretion flow (Kotov, Churazov \& Gilfanov 2001; Ar\'evalo \& Uttley 2006; Ingram \& van der Klis 2013).
These hard lags commonly show a decreasing trend of lag amplitude as a function of frequency that can be well described by a power law with spectral index $\sim-0.7$. It has been shown that in GX 339--4 during the \emph{hard state} the lag amplitude displays a much steeper decrease when the lags are computed with respect to a soft energy band containing significant contribution from the disc (Uttley et al. 2011; De Marco et al. 2015). This steepening is mostly due to a drop of lag amplitude at high frequencies, possibly associated with the intrinsic suppression of hard lags. At these frequencies a disc reverberation component (confirmed through the analysis of the lag-energy spectra) has been observed in archival data (Uttley et al. 2011; De Marco et al. 2015). Here we checked for this behaviour by fitting the lag-frequency spectra with a simple power law model over the frequency range 0.024-30 Hz. Results of the fits are reported in Table \ref{table3}.

\begin{table*}
	\centering
	\caption{The table reports relevant information derived from the analysis of X-ray lags: the best-fit values of the slope parameter obtained from fitting the \emph{very soft}-\emph{soft} ($\alpha_{vs-s}$, column 2) and the \emph{soft}-\emph{hard} ($\alpha_{s-h}$, column 3) lag-frequency spectra of each observation with a power law model over the frequency range 0.024-30 Hz;  the thermal reverberation lag amplitude ($\tau_{rev}$, column 4), as derived following the procedure described in Sect. \ref{sec:Ltrend}; the disc-to-Comptonized flux ratio ($f$, column 5, see Sect. \ref{sec:Ltrend}) as obtained from the fits, in the \emph{very soft} (0.3-0.7 keV) energy band, of the covariance spectra extracted in the frequency interval used for estimating the reverberation lag amplitude (reported within parentheses).}

			\label{table3}
	\begin{tabular}{ccccc} 
		\hline
		  (1)    &  (2)                     & (3)                     & (4) &(5)	\\
		  Obs & $\alpha_{vs-s}$  &   $\alpha_{s-h}$ & $\tau_{rev}$ (s) &$f$ \\
		\hline
		O1 &     -1.74$^{-2.17}_{+0.58}$  & -0.60$\pm 0.10$ & $0.008\pm0.003$  &$1.03\pm0.41$ [5--30 Hz] \\
		O2 &     -1.08$^{-0.29}_{+0.25}$   & -0.70$\pm 0.06$ & $0.014\pm0.006$  & $1.43\pm0.38$ [3--10 Hz]\\
		O3 &     -0.80$\pm 0.19$ & -0.80$\pm 0.05$ & $0.004\pm0.003$  & $2.27\pm0.83$ [1--10 Hz]\\
		O4 &     -0.89$^{-0.23}_{+0.22}$ & -0.78$\pm 0.08$ & $0.011\pm0.003$  &$2.50\pm0.70$ [1--5 Hz]\\
		O5 &     -0.90$^{-0.58}_{+0.42}$ & -0.78$\pm0.10$ & \multirow{2}{*}{$0.014\pm0.006$}  &\multirow{2}{*}{$3.79\pm1.77$ [0.3--1.5 Hz]} \\
		O6 &     -1.86$^{-3.23}_{+1.0}$ & -0.52$\pm 0.19$  & \\
		\hline
	\end{tabular}
\end{table*}

The measured lag-frequency spectra (Fig. \ref{fig:lagfreqO1O2}) show hard X-ray lags over almost all the sampled frequencies, in agreement with previous studies. However, the lag spectra between the \emph{very soft} and the \emph{soft} band are generally steeper than those between the \emph{soft} and \emph{hard} band (Table \ref{table3}). 
The lag profiles match each other well at low frequencies, but significant deviations are observed at $\nu \gsim 0.3\ {\rm Hz}$. At these frequencies the \emph{very soft}-\emph{soft} lags drop below the \emph{soft}-\emph{hard} lags (see also insets in Fig. \ref{fig:lagfreqO1O2} showing a zoom of this frequency interval on a linear y-axis). 
This means that, at high frequencies, the \emph{very soft}-\emph{soft} lags are systematically smaller than the typical hard lags associated with the primary X-ray continuum. As previously pointed out, this behaviour was already observed in GX 339--4 during its \emph{hard state} (Uttley et al. 2011; De Marco et al.  2015). 
In addition, during the {\it XMM} monitoring the \emph{very soft}-\emph{soft} lags at frequencies $\gsim 1-2\ {\rm Hz}$ are consistent with being negative (e.g. during O1, at $\nu>5\ {\rm Hz}$  the average lag amplitude is $\sim -0.007\pm 0.007\ {\rm s}$), where a negative delay corresponds to the \emph{very soft} photons lagging behind the \emph{soft} photons, as expected from disc reverberation.

To verify the presence of a reverberation lag at high frequencies where the drop in the \emph{very soft}-\emph{soft} lag-frequency spectra is observed, in Sect. \ref{sec:lagE} we studied the Fourier frequency-resolved energy spectra of the lags. 

\begin{figure}
	\includegraphics[width=0.93\columnwidth]{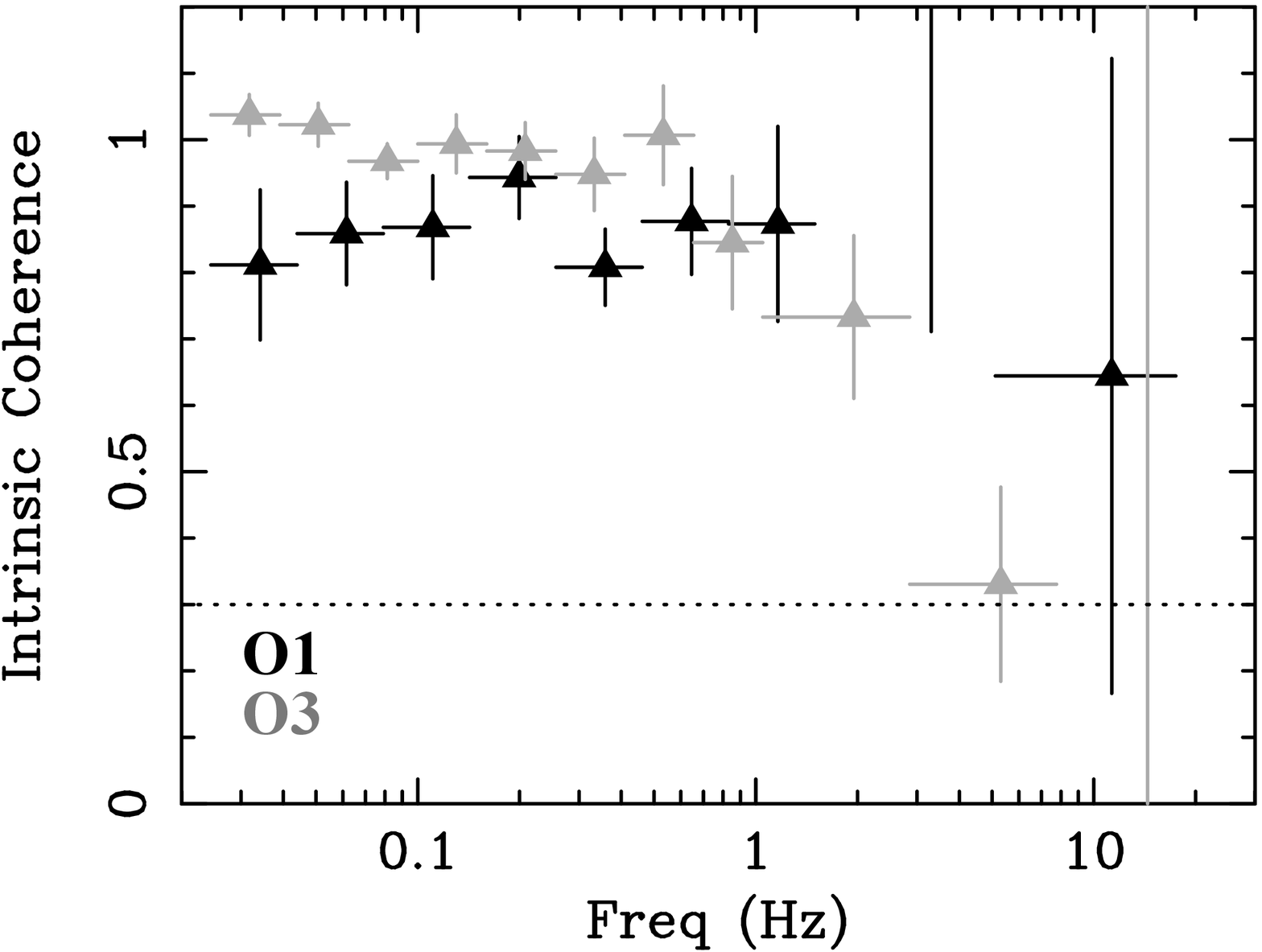}
	
    \caption{The intrinsic coherence as a function of frequency of O1 (black triangles) and O3 (gray triangles) between the {\emph very soft} and {\emph soft} energy bands. The dashed line indicates the threshold adopted for selection of the high frequency interval for the computation of the lag-energy spectra (see Sect. \ref{sec:lagE} for details).}
    \label{fig:coher}
\end{figure}

\begin{figure*}
	\includegraphics[width=0.9\textwidth]{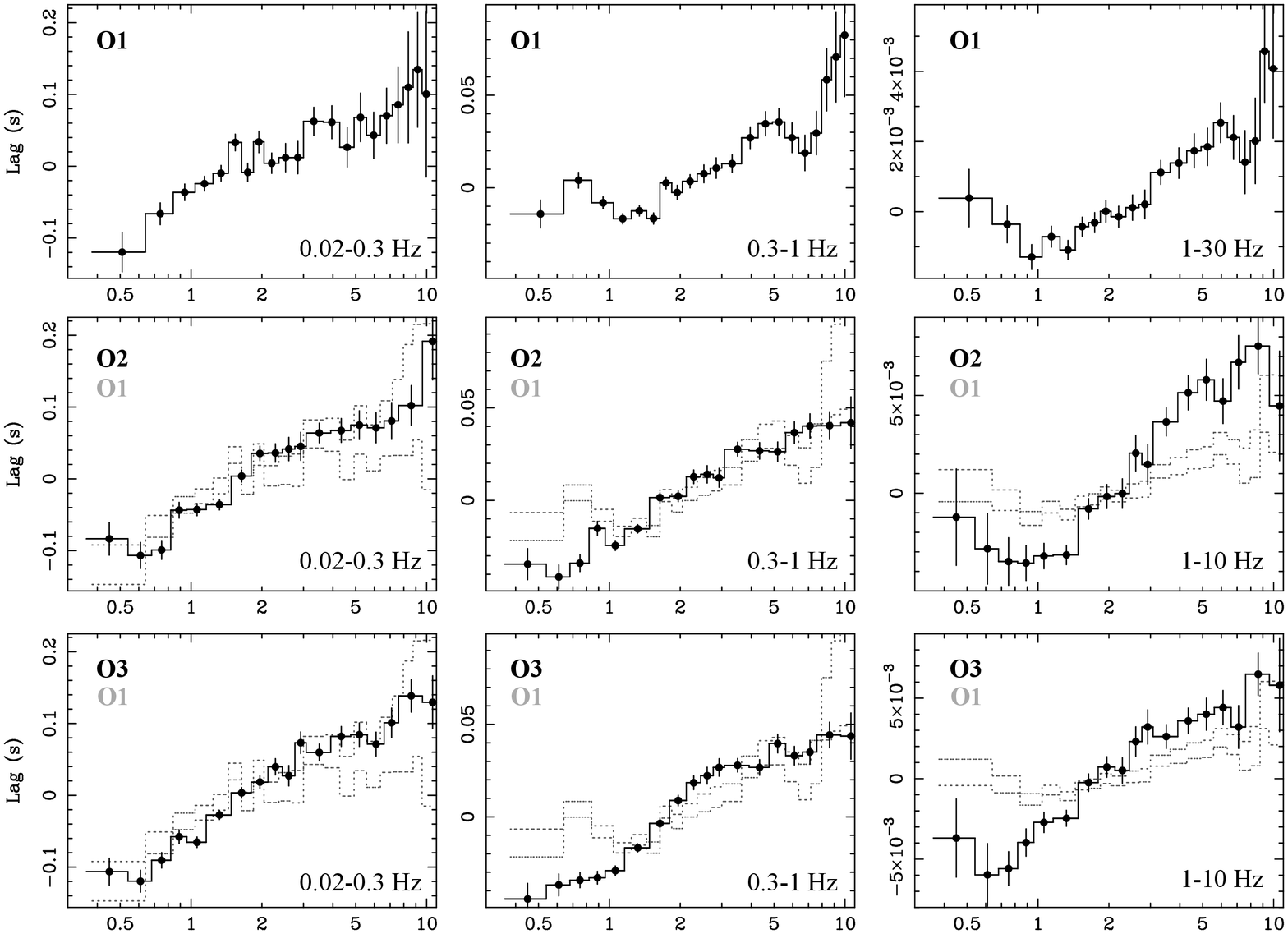}
	\includegraphics[width=0.9\textwidth]{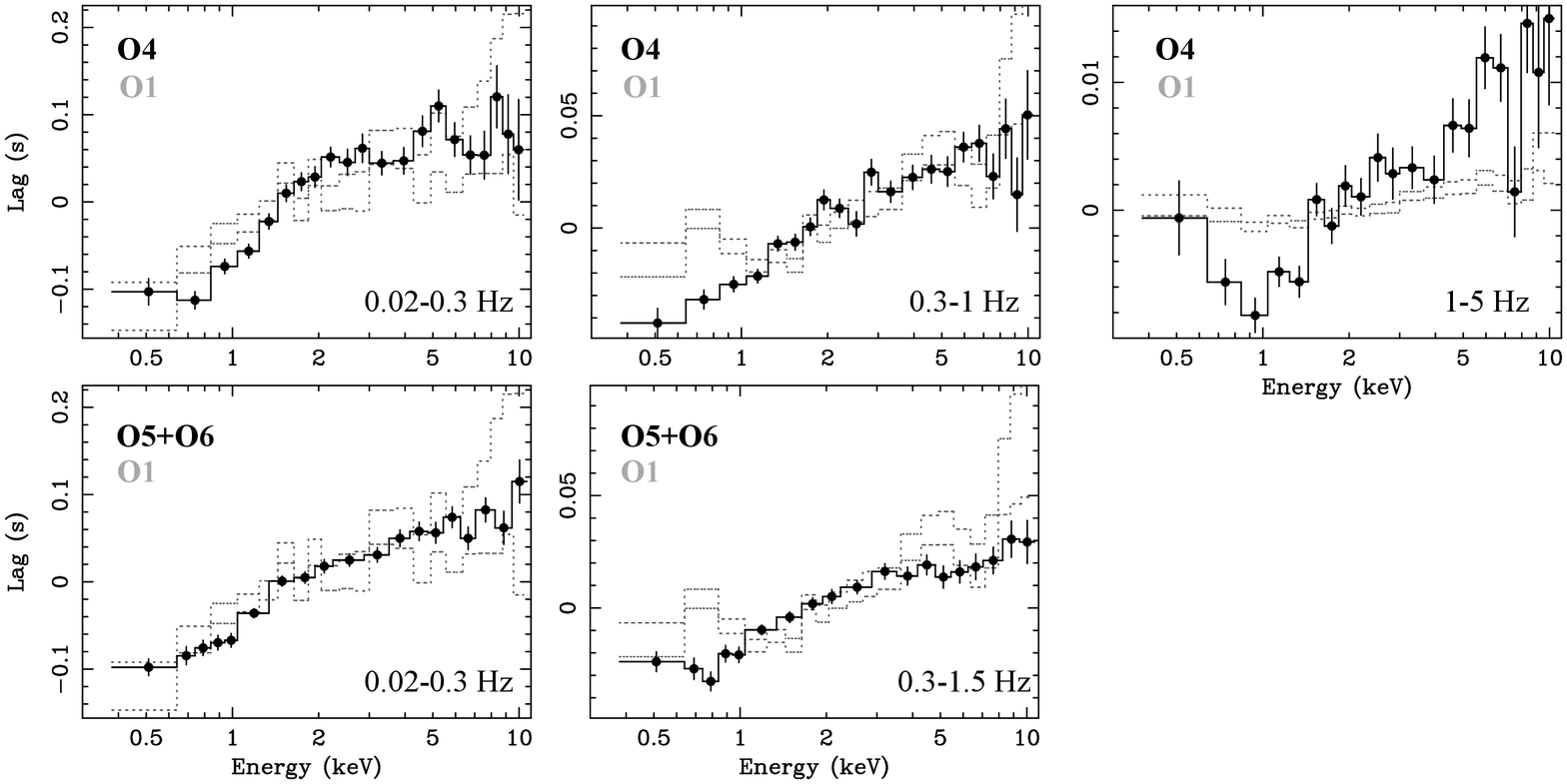}

    \caption{Lag-energy spectra of the different observations (O1 to O6, from top to bottom) in the \emph{low}, \emph{medium}, and \emph{high} frequency range (from left to right), as computed between a broad reference band (0.5-10 keV) and adjacent small energy bins. To facilitate comparison among the spectra of the different observations, for each frequency range we overplot the contours (gray dashed lines) of the lag-energy spectrum of O1.}
    \label{fig:lagEspecs}
\end{figure*}

 \subsection{Lag-energy spectra}
\label{sec:lagE} 

We computed the lag-energy spectra of GX 339--4, which measure the energy distribution of lag amplitude, within a given frequency interval. To this aim we followed the standard procedures extensively described in Uttley et al. (2014). We estimated lags between light curves extracted in a broad reference band (0.5-10 keV) and a series of adjacent energy bins (at each step, the counts from the energy bin are removed from the reference band light curve in order to avoid spurious contribution from correlated Poisson noise). The resulting cross-spectra were averaged over frequency intervals of interest, and the corresponding time lags plotted as a function of energy (Fig. \ref{fig:lagEspecs}).

The frequency intervals were chosen on the basis of the results of our analysis of X-ray lags as a function of frequency (Sect. \ref{sec:lagfreq}). We defined three relatively broad frequency intervals, continuously sampling the entire lag-frequency profile. The \emph{low frequency} interval (0.024-0.3 Hz) samples the range of frequencies where the \emph{very soft}-\emph{soft} and the \emph{soft}-\emph{hard} lag-frequency spectra match each other. The \emph{medium frequency} interval (0.3-1 Hz) samples the frequencies where the drop of \emph{very soft}-\emph{soft} lag amplitude starts to be observed. The \emph{high frequency} interval ($> 1\ {\rm Hz}$) samples the range of frequencies where our lag-frequency spectra analysis indicates the possible presence of negative amplitude \emph{very soft}-\emph{soft} lags. 
The highest frequency end of this interval depends on the specific data set and is limited by the presence of high-frequency uncorrelated noise. Indeed, we notice that the intrinsic coherence (computed using equation 8 of Vaughan \& Nowak 1997) between the very soft and soft light curves and between the soft and hard light curves is always high ($>$ 0.8) up to frequencies of a few-to-tens of Hz (depending on the observation, as an example we show in Fig. \ref{fig:coher} the intrinsic coherence of O1 and O3 between the \emph{very soft} and \emph{soft} bands), meaning that the light curves in the different bands have a high degree of linear correlation. At higher frequencies the intrinsic coherence becomes noisier (as a consequence of Poisson noise significantly affecting the data), and/or shows indications of a decline, particularly during observations with the lowest count rate. This decline might be either intrinsic or a consequence of the adopted correction for the raw coherence being inaccurate in this high frequency/low count rate regime. Therefore, to avoid spurious contributions from uncorrelated noise we chose the maximum frequency for our lag measurements by discarding those frequencies where the intrinsic coherence is either unconstrained or $\leq$ 0.3 (simulations of light curves contaminated with an uncorrelated signal show that this threshold allows measuring the lag with good accuracy, see Kara et al. 2013).
This maximum frequency ranges between 1.5 Hz (for the lowest count rate observation O6) and 30 Hz (in the case of the best quality O1 observation). The chosen frequency intervals are marked by coloured bars in Fig. \ref{fig:lagfreqO1O2} for O1, O2, and O3.
 
Lag-energy spectra were computed for each single observation. However, given the low count rate of the last observation, O6, we decided to combine it with O5 in order to increase the signal-to-noise of lag measurements. Moreover, since the high-frequency limit is $\sim$1.5 Hz in these observations, we merged the \emph{medium} and \emph{high frequency} intervals to obtain better quality lag spectra.

The resulting lag-energy spectra in each of the chosen frequency intervals are shown in Fig. \ref{fig:lagEspecs}. To allow comparison among the different observations, for each frequency range we overplot the contours of the corresponding lag-energy spectrum of O1 (gray dashed lines).

The main observed properties of the lag-energy spectra can be summarized as follows:
\begin{description}
\item[{\bf Low frequencies (0.024-0.3 Hz) -}] hard X-ray lags dominate the entire X-ray band 0.3-10 keV. The larger the separation between two energy bands the longer their relative delay. As observed in the past (e.g. Miyamoto \& Kitamoto 1989; Nowak et al. 1999), this trend can be broadly described by a log-linear model. In our case, a fit with a log-linear model yields best-fit slopes of $\sim0.1-0.2$. In this frequency range the lag-energy spectra of the different observations are all consistent with each other within their $1\sigma$ uncertainties.
\item[{\bf Medium frequencies (0.3-1 Hz) -}] the lag-energy spectra are more complex, significantly deviating from a log-linear trend. In general, hard lags still dominate at high energies ($E\gsim$ 2 keV). At these energies the lag-energy spectra do not show significant differences among the various observations. On the other hand, significant differences are observed in the soft band. These are due to the spectra breaking to a steeper profile at $E\lsim 2-3$ keV (particularly during O3 and O5$+$O6) and/or flattening out at $E\lsim$ 1 keV (during O1, O3, and O5$+$O6). During O5$+$O6 we observe indications of the emergence of a soft delay in the energy band dominated by the disc component.
\item[{\bf High frequencies ($>1$ Hz) -}] while the high energies are still dominated by hard lags, all the observations show the emergence of a soft lag at $E\lsim1$ keV, characterized by relatively small amplitudes (of the order of a few ms). This is indicative of a delayed response of soft photons. As shown in Sect. \ref{sec:specs}, the soft band contains significant contribution from disc emission (ranging between 23 and 57 percent of the total 0.3-1.5 keV flux, Fig. \ref{fig:ratios}). Therefore, these soft lags have the same properties as observed in previous analyses of \emph{hard state} observations of GX 339--4 (Uttley et al. 2011; De Marco et al. 2015) and H1743-322 (De Marco \& Ponti 2016), where the lags have been interpreted as the signature of thermal reverberation resulting from hard X-ray illumination of the disc. 
\end{description}

\begin{figure*}
	\includegraphics[width=\textwidth]{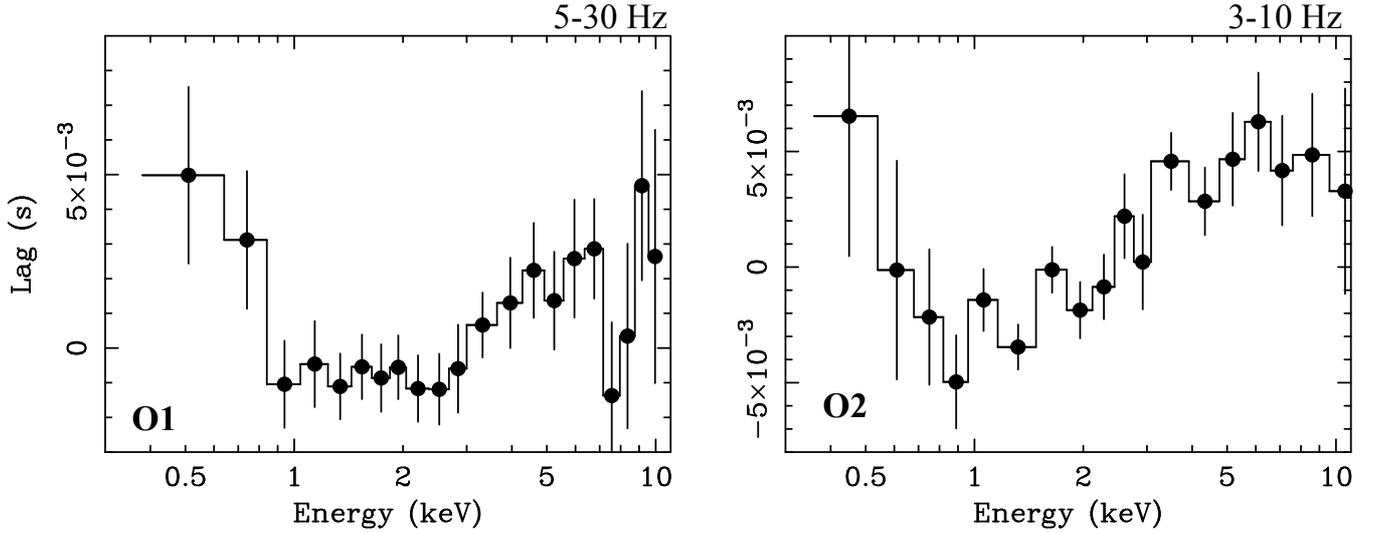}
    \caption{Lag-energy spectra of observations O1 (\emph{left panel}) and O2 (\emph{right panel}) restricted to the \emph{very-high frequency} range, which extends up to 30 Hz for O1 and up to 10 Hz for O2.}
    \label{fig:lagEspecsO1O2}
\end{figure*}

For the highest count rate, best quality observations O1 and O2, we refined our lag measurements, restricting the analysis to a {\bf very-high frequency} range, namely 5-30 Hz for O1 and 3-10 Hz for O2. The corresponding lag-energy spectra are shown in Fig. \ref{fig:lagEspecsO1O2}. As clear in the figure, at these frequencies the high-energy hard lags are less prominent and the soft lag stands out more clearly. This is expected in a thermal reverberation interpretation, in that the reverberation lag should be relatively small and span a wide range of frequencies. On the other hand the amplitude of the hard lags associated with the hard X-ray continuum decreases as the frequency increases (Sect. \ref{sec:lagfreq} and Fig. \ref{fig:lagfreqO1O2}), so that at sufficiently high frequencies the reverberation lag should be of the same order, or even dominate the lag spectra.

The highest flux observation (O1), which allows us to extend the analysis up to 30 Hz, shows signatures of additional complexities at energies $\gsim 1$ keV (see Fig. \ref{fig:lagEspecsO1O2}, left panel). Indeed, between $\sim1-3$ keV the relative lags are consistent with zero, indicating possible suppression of the hard lags associated with the primary X-ray continuum. On the other hand, the $E\sim 3-7$ keV energy band is delayed with respect to the $\sim1-3$ keV band by a few milliseconds. This suggests the presence of delayed emission from a separate spectral component dominating at $E\sim 3-7$ keV. 
Apart from the Comptonization continuum, the only component significantly contributing in this band is the Fe K emission line complex (Petrucci et al. in preparation). Therefore, the observed lag might be associated with the delayed response of the disc reflection component. We verified this hypothesis by fitting the lag-energy spectrum of O1 at $E>1$ keV with a log-linear model (which assumes the sole presence of hard lags associated with the hard X-ray continuum) and with a constant plus gaussian model (which assumes suppression of the hard lags associated with the hard X-ray continuum and the presence of FeK reverberation). The fits yield similar results ($\chi^2/dof \sim$ 11.65/17 and $\chi^2/dof \sim6.66/15$, respectively). Therefore, the quality of the spectrum does not allow us to prefer one of the two models. However, we notice that in the case of zero relative lags in the continuum, the significance of the $E\sim3-7$ keV component is assessed at $\sim3\sigma$. Though the feature appears broad, its width is not constrained.  
The fit with a gaussian yields a 99 percent lower limit on the width of $\sigma\gsim 0.2\ {\rm keV}$ (consistent with constraints on the Fe K line width as obtained from the fit of the energy spectrum; Petrucci et al. in preparation), and does not allow us to derive strong conclusions on the disc truncation radius. 
Interestingly, in this case the lags measured at the Fe K line centroid energy are of the same amplitude as the soft lag ascribable to thermal reverberation. This is consistent with the two components being produced in the same region of the disc.

\begin{figure}
	\includegraphics[width=\columnwidth]{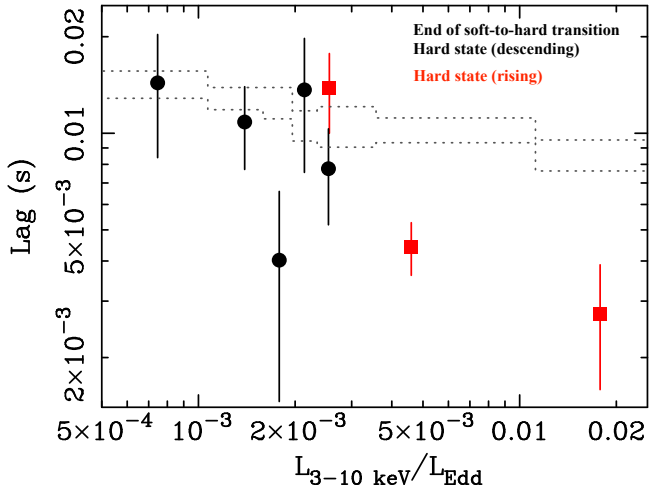}
    \caption{The reverberation lag amplitude as a function of 3-10 keV Eddington-scaled luminosity. The black dots refer to the estimates obtained in this paper. These data cover the end of the transition from the \emph{soft to the hard state}, and the decrease of luminosity through the \emph{hard state} preceding the return to quiescence at the end of the 2014-2015 outburst. The red squares are estimates obtained in De Marco et al. (2015) from the analysis of archival {\it XMM} observations covering the rise of luminosity through the \emph{hard state} during the 2009 and the 2004 outburst. The gray dotted lines represent the $1\sigma$ contours showing the expected trend of observed lag amplitude as a function of luminosity induced by variations of the relative fraction, $f$, of thermally reprocessed flux to the direct variable continuum, and assuming a constant value for the intrinsic reverberation lag (see Sect. \ref{sec:Ltrend}).}
    \label{fig:LagvsL}
\end{figure}

\section{Is the soft lag due to thermal reverberation?}
\label{sec:results}
Our systematic study of GX 339--4 at the end of the transition from the \emph{soft to the hard state} and during the descending phase in the \emph{hard state} revealed a strong dependence of the X-ray lags on the sampled variability time scale. On time scales of tens to a few seconds (corresponding to the \emph{low frequency} interval, Sect. \ref{sec:lagE}) hard X-ray lags dominate the lag-energy spectra. All the low-frequency lag-energy spectra can be consistently modeled with a log-linear model, with a best-fit slope of $\sim$0.1-0.2.
 
On time scales of a few-to-fractions of seconds (corresponding to the \emph{high frequency} interval, Sect. \ref{sec:lagE}) we observe the appearance of a soft lag at $E\lsim1$ keV, where the disc component gives a significant contribution (23 and 57 percent of the total flux). The measured lag amplitude is of the order of a few milliseconds. At sufficiently high frequencies this lag is more prominent as a consequence of the hard lag amplitude decreasing steadily as a function of frequency (Sect. \ref{sec:lagfreq} and Fig. \ref{fig:lagfreqO1O2}). We showed this for the two best quality observations, O1 and O2 (Fig. \ref{fig:lagEspecsO1O2}), considering a \emph{very high frequency} window which samples time scales of $\sim$0.2-0.02 s. For the other, lower count rate observations, our ability to extend the analysis to such high frequencies is limited by the presence of a dominant uncorrelated noise component (Sect. \ref{sec:lagE}).

The analogies with the soft lag detected in archival data of GX 339--4 and H1743--322 in the \emph{hard state} (Uttley et al. 2011; De Marco et al. 2015; De Marco \& Ponti 2016) and ascribed to disc thermal reverberation suggest a common origin. 
Here we test this hypothesis by investigating variations of the lag amplitude as a function of the accretion regime.

\subsection{Trends of lag amplitude as a function of luminosity}
\label{sec:Ltrend}
Reverberation lags due to reprocessing of primary hard X-ray photons in the disc are expected to map the variations of disc-hot flow geometry thought to occur during an outburst (Sect. \ref{sec:intro}). 
In our previous studies (De Marco et al. 2015; De Marco \& Ponti 2016) we observed a dependence of the soft lag amplitude on the source luminosity. This is in agreement with the lag being produced by thermal reverberation and the observed trend due to variations of inner flow geometry in the \emph{hard state} at the beginning of the outburst. The new data set allows us to extend this analysis to the end of the \emph{soft-to-hard state} transition and the return to quiescence.

As discussed in De Marco et al. (2015) and De Marco \& Ponti (2016), obtaining an estimate of the amplitude of the soft lag ascribable to thermal reverberation requires disentangling it from the underlying hard lags associated with the X-ray continuum. Here, for simplicity and for consistency with our previous works we followed the simple approach of estimating the maximum intensity of the residuals, in the \emph{very soft} band, above the extrapolation of the log-linear model best-fitting the high energy hard lags. 
We applied this procedure to the lag-energy spectra in the frequency interval where the soft lag is observed (i.e. the \emph{very high frequency} interval for O1 and O2, Fig. \ref{fig:lagEspecsO1O2}, the \emph{high frequency} interval for O3 and O4, and the \emph{medium frequency} interval extended to 1.5 Hz for O5+O6, Fig. \ref{fig:lagEspecs}).

The hard lags have been fit with a log-linear model at $E>0.8$ keV. 
It is worth noting that, while this model provides a good description of the overall energy dependence of the hard lags in BHXRBs, recent works have highlighted deviations from a single-slope log-linear model when the analysis is extended to the band containing contribution from the disc (Uttley et al. 2011; Cassatella, Uttley \& Maccarone 2012; De Marco et al. 2015). In particular, in the analysed data set we observe the presence of a break to a steeper slope at $E\lsim 2-3$ keV during some observations, specifically O3 and O5$+$O6. For these observations we limited our fit with a single log-linear model to the energy range E$\sim 0.8-3$ keV.

The resulting estimates of soft lag amplitude range between $\sim 0.004-0.014$ s (Table \ref{table3}). These values are plotted in Fig. \ref{fig:LagvsL} (black points) as a function of 3-10 keV Eddington scaled luminosities. 
In order to compare these results with previous detections we considered the lag measurements obtained from our previous analysis of archival {\it XMM} observations of GX 339--4 in the \emph{hard state} (De Marco et al. 2015; overplotted as red squares in Fig. \ref{fig:LagvsL}). This comparison highlights a net decrease of soft lag amplitude (by a factor $\sim$5) as a function of luminosity, as expected if produced by thermal reverberation, therefore in agreement with previous interpretations (De Marco et al. 2015; De Marco \& Ponti 2016). Fitting these data with a linear model in log-space we obtain a best-fit slope parameter of $-0.56\pm0.28$.

While the observed variations of lag amplitude as a function of luminosity can be the result of variations of intrinsic reverberation lag amplitude, variations of the strength of the thermally reprocessed component relative to the direct, variable continuum might also produce a similar trend. 
This can be easily seen by assuming that the reference band mostly contains primary hard X-ray photons while the \emph{very soft} X-ray band contains both primary photons and reprocessed disc emission. In this case, for small phase lags (where the phase, $\phi$, and time lag, $\tau$, are related by $\tau=\phi/ 2\pi \nu$) and assuming that the hard lags intrinsic to the primary continuum are approximately equal to zero at high frequencies:
$$ 
\phi_{obs}\approx \phi_{i} \frac{f}{1+f}
$$
where, $\phi_{obs}$ and $\phi_{i}$ are, respectively, the measured and intrinsic reverberation phase lag, while $f$ is the ratio between the reprocessed flux responsible for the reverberation lag and the variable direct continuum. 
Using this formula we verified the hypothesis that the observed trend of lag amplitude with luminosity is driven by variations of reverberation lag intrinsic amplitude rather than variations of $f$. To this aim we estimated $f$ from the covariance spectra (Wilkinson \& Uttley 2009; Uttley et al. 2014), which return the flux distribution of linearly correlated spectral components. The covariance spectra were computed in the same frequency intervals as used for estimating the reverberation lag amplitude. Indeed, at these frequencies the dominant contribution to the disc component in the covariance spectra should come from thermally reprocessed photons rather than intrinsic, variable disc emission. We fit the covariance spectra with the same underlying model used in Sect. \ref{sec:specs} for the fit of the time-averaged energy spectra, and used the disc-to-Comptonized flux ratios from the covariance spectra in the \emph{very soft} band as an estimate of $f$. The derived values of $f$ for the observations of the {\it XMM} monitoring are reported in Table \ref{table3}, while those obtained from the archival {\it XMM} observations are $f=0.91\pm0.19$, $f=1.32\pm0.27$, and $f=2.01\pm0.64$, respectively for the highest, medium, and lowest sampled luminosities in the \emph{hard state}. 
Finally, we assumed a constant value for the intrinsic reverberation lag and measured the variations of observed lag amplitude induced by the observed variations of $f$. The results are plotted in Fig. \ref{fig:LagvsL} (the dotted gray lines are obtained considering the $1\sigma$ error on $f$). We infer that variations in the relative fraction of reprocessed-to-direct flux would produce a decrease of observed lag amplitude of a factor $\lsim 2$ as a function of luminosity, significantly smaller than the measured decrease of a factor $\sim 5$. Assuming a normal distribution for the reverberation lag estimates in Fig. \ref{fig:LagvsL}, we note that the decrease observed in the data is such that the two points at the highest luminosities deviate at $\sim 5-7\sigma$ from the trend expected assuming a constant lag and varying $f$. However, assessing the significance of these deviations via more rigorous statistical tests is hampered by the limited number of data points.
 Therefore, we conclude that the observed trend is consistent with being driven by variations of intrinsic reverberation lag amplitude, likely resulting from variations of inner flow geometry.

\section{Discussion}
\label{sec:discussion}
We carried out a spectral-timing study of GX 339--4 during its last outburst (started in 2014 October and ended in 2015). The analysed observations cover the end of the transition from the \emph{soft to the hard state} and the decrease of luminosity in the \emph{hard state} preceding the return to quiescence (Sect. \ref{sec:state}). The 3-10 keV luminosity of GX 339--4 during the first four observations (O1-to-O4) is in the range $\sim0.0025-0.0014\  \rm{L_{Edd}}$. Assuming the best-fit model described in Sect. \ref{sec:specs}, these values correspond to a bolometric luminosity of $\sim0.01\  \rm{L_{Edd}}$ and are in agreement with estimates of \emph{soft-to-hard} transition luminosities obtained from BHXRB population studies (e.g. Maccarone 2003; Dunn et al. 2010). The luminosity of GX 339--4 then decreases quickly during the last two observations (O5 and O6), by a factor of $\sim$3 and 5 over a time span respectively of 6 and 18 days.

Our spectral study revealed the ubiquitous presence of emission at soft X-ray energies in excess of the hard X-ray Comptonization component (Sect. \ref{sec:specs}). This ``soft X-ray excess'' has been observed in several \emph{hard state} observations of both GX 339--4 and other BHXRBs (e.g. Takahashi et al. 2008; Cassatella et al. 2012; Kolehmainen et al. 2014). We modeled this component as thermal emission from the disc, and measured a decrease of its relative contribution from $\sim$57 (during the first observation) to 23 percent (during the last observation) of the total 0.3-1.5 keV flux (Fig. \ref{fig:ratios}).
Given the resulting peak temperature ($\sim$0.2 keV), this component can be ascribed to emission from a truncated and/or low luminosity disc.
 
Full radiative coupling between the cold disc and the hot Comptonizing medium implies a fraction of Comptonized hard X-ray photons to be reprocessed in the disc. Part of these photons are reflected (e.g. George \& Fabian 1991), while the non-reflected photons are reprocessed and re-emitted as quasi-thermal emission (e.g. Malzac, Dumont \& Mouchet 2005; Gierli\'nski, Done \& Page 2008). The latter component adds on the intrinsic disc emission and is expected to contribute significantly to the soft X-ray spectrum whenever the intrinsic disc emission is weak, such as during \emph{hard} and \emph{hard-intermediate states}. Therefore we infer that the ``soft X-ray excess'' detected during the observations of GX 339--4 analysed in this paper likely includes a contribution from thermally reprocessed emission.

Our study of X-ray lags in GX 339--4 supports this interpretation, in that the thermally reprocessed emission is expected to directly respond to variations of the hard X-ray illuminating continuum, with a time delay mostly due to the light travel time between the Comptonization region and the reprocessing region in the disc (Uttley et al. 2011; 2014).
Indeed, we observe that the disc thermal emission responds to the fast variability (on time scales of a few-to-fractions of seconds) of the hard X-ray flux with a time delay in the range $\sim$0.004-0.014 s (Sect. \ref{sec:lagE}, Table \ref{table3}, and Fig. \ref{fig:LagvsL}), and we interpret this delay as a signature of disc thermal reverberation. 
The lack of QPOs in the power spectra of GX 339--4 (Sect. \ref{sec:state}) leads us to the conclusion that the detected lags are associated with the broad band noise continuum rather than related to the complex phenomenology characterizing QPOs (e.g. Casella et al. 2005; Stevens \& Uttley 2016; Ingram et al. 2016; van den Eijnden et al. 2017).

In addition to the soft, thermal reverberation lag, we find indications of a reverberation lag at the energies of the Fe K line complex (Sect. \ref{sec:lagE} and Fig. \ref{fig:lagEspecsO1O2} left panel). Interestingly, the Fe line shows a time delay of similar amplitude as that of the thermally reprocessed component. This meets the expectations of the disc reverberation interpretation, as the same irradiating photons are responsible for the production of both the reflected and the thermalized components. 
However, this detection needs to be confirmed through longer observations and/or broader energy coverage. 

During the monitoring and within the uncertainties associated with our lag estimates, the reverberation lag amplitude does not show strong variations. However, comparing these results with those we obtained from the study of the disc thermal reverberation in GX 339--4 during the rise of the \emph{hard state} at the beginning of the outburst (De Marco et al. 2015, red squares in Fig. \ref{fig:LagvsL}), a decrease of reverberation lag amplitude as a function of luminosity by a factor of $\sim5$ is observed (Sect. \ref{sec:Ltrend} and Fig. \ref{fig:LagvsL}). 
This behaviour is in line with the scenario depicted by truncated disc models (e.g. Esin et al. 1997), in that the inner disc radius should move inward as the source luminosity increases in the \emph{hard state}, and should recede during the transition from the \emph{soft} to the \emph{hard state} and the return to quiescence.

The average amplitude of the reverberation lag during the monitoring is 0.009$\pm$0.002 s. This value is consistent with that measured in the \emph{hard state} at similar X-ray luminosities, but during the initial phases of the outburst (lowest luminosity observation in De Marco et al. 2015; see also Fig. \ref{fig:LagvsL}).
This result indicates that, at a given luminosity during \emph{hard-intermediate} and \emph{hard states}, the inner accretion flow geometry is similar from outburst to outburst of the same source, independently as to whether the source is in the initial or final phases of an outburst. It is worth stressing that this inference applies to accretion states characterized by relatively high spectral hardness (we measure $\rm{F_{6-10 keV}/F_{3-6 keV}}\gsim$0.8 both during the archival observations and during the {\it XMM} monitoring; see Fig. \ref{fig:HID}). On the other hand, independent estimates of the inner disc radius via spectroscopic methods (e.g. Steiner et al. 2010; Plant et al. 2014) support the presence of a standard disc reaching the last stable orbit during softer states. Therefore, a dependence of reverberation lag amplitude on the spectral hardness is also expected. We could not find any significant evidence of such a dependence in the analysed data set, most likely because of the relatively small range of hardness ratios spanned by the source.  
Nonetheless, even though reverberation lag measurements have not yet been performed during disc-dominated states, we expect the amplitude to be smaller by a factor 10-100 than measured here.

The measured reverberation lag amplitudes can be translated into gravitational radii light crossing time units ($r_g/c$), to gain an idea of the distance between the source of hard X-rays and the reprocessing region in the disc. Assuming a low inclination disc (e.g. as inferred from the non-detection of equatorial winds in the \emph{soft state}, Ponti et al. 2012a, 2016, and from the shape of the HID, Mu\~noz-Darias et al. 2013), a black hole mass of $12\ \rm{M}_{\odot}$ (Sect. \ref{sec:state}), and a central source illuminating a truncated, face-on disc, the measured lags correspond to distances of the order of $\sim 40\ r_g$ (at high luminosity, for a lag of $\sim 0.003\ {\rm s}$) to $\sim 200\ r_g$ (at low luminosity, for a lag of $\sim 0.014\ {\rm s}$). 
As a consistency check we derived the expected inner temperature, $kT_{d}$, of a truncated disc, using the values of the truncation radius derived from reverberation measurements and estimating the disc bolometric luminosity from the best-fit models presented in Sect. \ref{sec:specs} (e.g. Kubota et al. 1998). The derived values of the disc inner temperature range between $kT_{d}\sim$0.07 keV (at large truncation) and $kT_{d}\sim$0.19 keV (at low truncation). Assuming a color temperature correction factor of $\sim1.9-2.2$ (equation A13 in Davis, Done, \& Blaes 2006) these values are close to those obtained from our spectral fits (Table \ref{table2} and De Marco et al. 2015 for the archival {\it XMM} observations). 
Of course these must be taken as back-of-the-envelope estimates, since more precise estimates of the disc inner radius require accounting for several observational aspects (as discussed in De Marco \& Ponti 2016) and the use of self-consistent spectral timing models.

It is worth noting that the main argument against truncated disc models comes from the detection, in the \emph{hard state}, of a broad component of the Fe K emission line. 
Though several studies ascribe the observed broadening to reflection in a truncated and ionized disc (e.g. Plant et al. 2015; Basak \& Zdziarski 2016), this might as well be explained as due to relativistic effects (e.g. Fabian et al. 1989; Done et al. 2007) occurring in a disc extending close to the last stable orbit (e.g. Miller et al. 2004; Miller et al. 2006a, 2006b; Garc\'ia et al. 2015). 
If the disc is non-truncated during the observations analysed in this paper, then the observed reverberation lags require the source of hard X-ray photons to illuminate the inner disc from a height ranging between a few tens to $\sim 100\ r_g$ above the disc.
For example, the production of hard X-ray photons in a collimated structure, such as a jet (e.g. Fabian et al. 2014), might explain the longest lags, although collimation would also reduce the fraction of hard X-ray photons illuminating the disc. Nonetheless, even in a non-truncated disc scenario, the observed variations of lag amplitude as a function of luminosity would imply variations of the geometry of the source of hard X-ray photons, moving closer to the disc or becoming more compact as the luminosity increases.

\section{Conclusions}

We have presented a systematic study of X-ray reverberation lags at the end of the 2014-2015 outburst of the BHXRB GX 339--4. The analysed observations caught the last phases of the \emph{soft-to-hard state} transition and the decrease of luminosity in the \emph{hard state} before the return to quiescence. 
Our main conclusions can be summarized as follows:
\begin{enumerate}
\item A soft X-ray lag ascribable to disc thermal reverberation is observed during all the observations when frequencies $\gsim$ 1 Hz are sampled. Given the absence of QPOs, the lag is associated with the broad band noise variability components.
\item Comparing the new detections with those obtained from the analysis of archival data sets (De Marco et al. 2015) we observe a net decrease of lag amplitude (by a factor $\sim$ 5) as a function of luminosity, confirming our previous analyses (De Marco et al. 2015 and De Marco \& Ponti 2016). 
\item We ascribe the observed dependence of reverberation lag amplitude on luminosity to variations of the geometry of the inner accretion flow. In particular, an inner disc truncation radius approaching the ISCO as the luminosity increases at the beginning of the outburst and receding as the luminosity decreases at the end of the outburst is in agreement with our results. Alternative explanations might involve a disc always reaching ISCO and the source of hard X-ray photons moving closer to the disc/becoming more compact as the luminosity increases.
\item During \emph{hard-intermediate} and \emph{hard states}, when similar luminosities are sampled, the reverberation lag amplitude does not change significantly from outburst to outburst, suggesting similar inner flow geometry.
\item We found hints of reverberation in the FeK component during the first observation (O1). The lag measured at the line centroid is consistent with the disc reprocessing interpretation, in that it has the same amplitude as the lag associated with the thermal reverberation component, suggesting the two are produced in the same region of the disc. Longer observations and/or broader energy coverage are needed to confirm this detection.
\end{enumerate}

\section*{Acknowledgements}
This work is part of the CHAOS project ANR-12-BS05-0009 supported by the French Research National Agency
(http://www.chaos-project.fr). The authors acknowledge also funding support from the CNES and the PNHE.
The analysis is based on observations obtained with {\it XMM-Newton}, an ESA science mission with instruments and contributions directly funded by ESA Member States and NASA. 
The authors thank the referee for the helpful comments.
BDM acknowledges support from the European Union's Horizon 2020 research and innovation programme under the 
Marie Sk{\l}odowska-Curie grant agreement No. 665778 via the Polish National Science Center grant Polonez 
UMO-2016/21/P/ST9/04025. AAZ acknowledges support from Polish National Science Centre grants 2013/10/M/ST9/00729 and 2015/18/A/ST9/00746.












\bsp	
\label{lastpage}
\end{document}